\documentclass[fleqn,twoside]{article}
\usepackage{espcrc2}
\usepackage{amsmath}
\def\lsim{\raise0.3ex\hbox{$<$\kern-0.75em\raise-1.1ex\hbox{$\sim$}}}
\def\gsim{\raise0.3ex\hbox{$>$\kern-0.75em\raise-1.1ex\hbox{$\sim$}}}

\usepackage{graphicx}
\usepackage[figuresright]{rotating}

\title{Saturation in DIS at low x}

\author{D. Schildknecht\address[DPUB]{Department of Physics, 
        University of Bielefeld, \\ 
        P.O. Box 10 01 31, 33501 Bielefeld, Germany}
        \thanks{Supported by DFG, contract Schi 189/6-2}
        \thanks{Presented at Diffraction
         2004, Cala Gonone, Italy, September 18-23, 2004}}
       
\begin{document}

\begin{abstract}
Saturation at low x appears as an almost unavoidable consequence of the
two-gluon exchange generic structure.
\vspace{1pc}
\end{abstract}

\maketitle


In this written version of my talk I will restrict myself to giving a
brief discussion on the empirical evidence for the concept of ``saturation''
at low x in deep inelastic lepton-nucleon scattering.

In the model-independent analysis of the experimental data from HERA on DIS
at low x carried out in the summer of the year 2000, 
we found\cite{Diff2000}
that the data
on the total virtual photoabsorption cross section lie on a universal curve
when plotted against the dimensionless variable
\vspace*{-0.2cm}
\begin{equation}
\eta = \frac{Q^2 + m^2_0}{\Lambda^2_{sat} (W^2)},\label{(1)}
\end{equation}\vspace*{-0.2cm}
where\vspace*{-0.2cm}
\begin{equation}
\Lambda^2_{sat} (W^2) = B \left( \frac{W^2}{W^2_0} + 1 \right)^{C_2} \simeq B
\left( \frac{W^2}{W^2_0} \right)^{C_2}.\label{(2)}
\end{equation}
Compare fig. 1. The energy-dependent quantity, $\Lambda^2_{sat} (W^2)$, acts
as the scale (``saturation scale'' or ``saturation momentum'') that determines
the range of $Q^2$ in which the energy dependence (at fixed $Q^2$) is
either hard $(\eta >> 1)$ or soft $(\eta << 1)$. The model-independent analysis
only rest on the assumption that $\sigma_{\gamma^*p} (W^2, Q^2)$ be a smooth
function of $\eta$. The fitting procedure gave\cite{Diff2000,Cvetic}
\vspace*{-0.1cm}
\begin{eqnarray}
m^2_0 & = & 0.15 \pm 0.04 GeV^2,\nonumber \\
W^2_0 & = & 1081 \pm 12 GeV^2, \nonumber \\
C_2 & = & 0.27 \pm 0.01.\label{(3)}
\end{eqnarray}
As long as only smoothness of $\sigma_{\gamma^*p}$ is assumed, the constant
$B$ can be arbitrary. With the explicit form of $\sigma_{\gamma^*p}$ in the
generalized vector dominance-color dipole picture (GVD-CDP), we found
\vspace*{-0.1cm}
\begin{equation}
B = 2.24 \pm 0.43 GeV^2.\label{(4)}
\end{equation}

\begin{figure}[htb]
\vspace{9pt}
\includegraphics{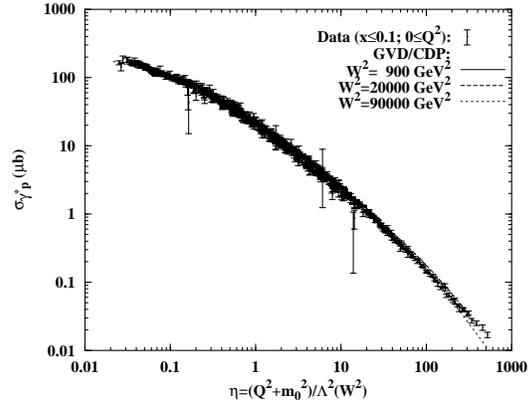}
\vspace*{3cm}
\caption{The total photoabsorption cross section as a function of the scaling
variable $\eta$ from (1).}
\vspace*{-0.7cm}
\label{fig1}
\end{figure}

\begin{figure}[thb]
\vspace*{6.0cm}
\includegraphics{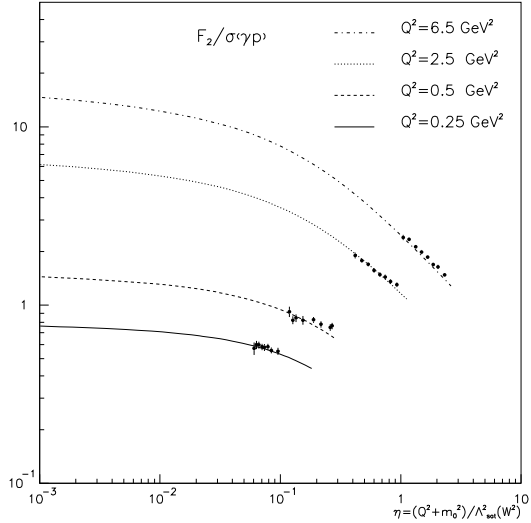}
\vspace*{0.3cm}
\caption{The ratio of the structure function $F_2 (x, Q^2)$ and the
photoabsorption cross section as a function of $\eta$.}
\label{fig2}
\vspace*{-0.7cm}
\end{figure}
Note that the data shown in fig. 1 include all data available for $x \simeq
Q^2/W^2 < 0.1$ and $0 \le Q^2 < 1000 GeV^2$, in particular, photoproduction
$(Q^2 = 0)$ is included.

Since the HERA energy, $W$, is limited, for large values of $Q^2$ small values
of $\eta << 1$ cannot be explored. The low-$\eta$ region in fig. 1 contains
data close to photoproduction, while the large-$\eta$ region is populated by
large-$Q^2$ measurements. Nevertheless, fig. 1 suggests that the ``saturation''
property\cite{Diff2000,Cvetic}
\vspace*{-0.1cm}
\begin{equation}
\lim_{{W^2 \to \infty}\atop{Q^2 fixed}} \frac{\sigma_{\gamma^*p} 
(\eta (W^2, Q^2))}{\sigma_{\gamma p} (W^2)} = 1\label{(5)}
\end{equation}
to be valid for any fixed $Q^2$.

In terms of the structure function
\vspace*{-0.1cm}
\begin{equation}
F_2 (x, Q^2) \simeq \frac{Q^2}{4 \pi^2 \alpha} \sigma_{\gamma^*p} (\eta
(W^2, Q^2)),\label{(6)}
\end{equation}
where $x \simeq Q^2/W^2$, according to (5) we have
\vspace*{-0.1cm}
\begin{equation}
\lim_{{W^2 \to \infty}\atop{Q^2 fixed}} 4 \pi^2 \alpha 
\frac{F_2 (x, Q^2)}{\sigma_{\gamma p}(W^2)} = Q^2.\label{(7)}
\end{equation}
An explicit empirical test of the approach to saturation accordingly requires
to plot the data for the ratio of the structure function $F_2 (x, Q^2)$ and
the photoproduction cross section as a function of $\eta$ at fixed $Q^2$.
Saturation requires the ratio (7) to become flat and approach the value
of $Q^2$ as a function of $\eta$ as soon as $\eta$ becomes small,
$\eta << 1$. 

The plot of the experimental data in fig. 2\cite{Schi-Ku}, 
for $Q^2 \lsim 0.5 GeV^2$ shows
the expected flattening in the $\eta$-dependence for $\eta << 1$. For larger
values of $Q^2$ the expected flattening for $\eta \lsim 0.1$ cannot be 
verified at present due to lack of energy.

No explicit theoretical ansatz is needed for the plots in figs. 1 and 2.
We have nevertheless included the theoretical curves 
from the GVD-CDP\cite{Diff2000,Cvetic,Ku-Schi} that
provides a theoretical basis for the observed scaling in 
$\eta$.\footnote{During discussion, I was asked which variable ``from the
point of view of the consumer should be bought'', $\eta$ or $\tau \sim
(x/x_0)^\lambda$\cite{Golec}. The answer: both, experiment is the arbiter.
The choice of $\Lambda^2_{sat} (W^2)$ follows from the mass dispersion 
relation of generalized vector dominance and as such is well-motivated.}

As conjectured\cite{Sakurai,Fraas} a long time ago, 
DIS at low x in terms of the 
virtual-photon-proton Compton amplitude is to be understood in terms of 
diffractive forward scattering of the hadronic $(q \bar q)^{J=1}$ (vector) 
states the virtual photon dissociates or fluctuates into. With the advent
of QCD, the underlying Pomeron exchange became understood in terms of
the coupling of two gluons\cite{Low} to the $(q \bar q)^{J=1}$ state. The
gauge-theory structure implies that the $(q \bar q)^{J=1}_{T,L}~p$
color-dipole cross section, proportional to the imaginary part of the 
$(q \bar q)^{J=1}_{T,L}~p$ forward-scattering amplitude, 
takes the form\cite{Nikolaev,Ku-Schi}
\vspace*{-0.2cm}
\begin{eqnarray}
&&\hspace*{-0.8cm} \sigma_{(q \bar q)^{J=1}_{T,L} p} 
(\vec r_\perp^{~\prime}, W^2)  \int
d^2 \vec l^{~\prime}_\perp \bar \sigma_{(q \bar q)^{J=1}_{T,L}p} 
(\vec l^{~\prime 2}_\perp, W^2) \cdot \nonumber \\
&& \cdot (1 - e^{-i\vec l^{~\prime}_\perp \vec r^{~\prime}_\perp})\\
& \simeq & \sigma^{(\infty)} \begin{cases}
1, & $$ {\rm for}~\vec r^{~\prime 2}_\perp \to \infty,$$ \nonumber \\
\frac{1}{4} \vec r^{~\prime 2}_\perp \Lambda^2_{sat} (W^2), & $${\rm for}~
\vec r^{~\prime 2}_\perp \to 0,$$
\end{cases} \label{(8)}
\end{eqnarray}
where by definition
\begin{equation}
\sigma^{(\infty)} \equiv \pi \int d \vec l^{~\prime 2}_\perp 
\bar \sigma_{(q \bar q)^{J=1}_L} (\vec l^{~\prime 2}_\perp, W^2), \label{(9)}
\end{equation}
and 
\vspace*{-0.2cm}
\begin{equation}
\Lambda^2_{sat} (W^2) \equiv \frac{\pi}{\sigma^{(\infty)}} 
\int d l^{~\prime 2}_\perp \vec l^{~\prime 2}_\perp 
\bar \sigma_{(q \bar q)^{J=1}_L} (\vec l^{~\prime 2}_\perp, W^2). \label{(10)}
\end{equation}
The (virtual) photoabsorption cross section is obtained from (8) by 
multiplication with the (light-cone) photon-wave function and subsequent
integration over the transverse  $q \bar q$ separation 
$\vec r_\perp = \vec r_\perp^{~\prime}/ \sqrt{z (1-z)}$ and the variable 
$z$ with $0 \le z
\le 1$ that e.g. determines angular distribution of the quark in the
$q \bar q$ rest frame.

It is important to note that the two-gluon-exchange dynamical mechanism 
evaluated for $x \to o$ implies the existence  of the saturation scale
$\Lambda^2_{sat} (W^2)$ according to (8). The scale $\Lambda^2_{sat}
(W^2)$ is related to the effective value of the gluon transverse momentum,
$\vec l_\perp = \vec l^{~\prime}_\perp \sqrt{z(1-z)}$, that enters the 
photoabsorption cross section as a consequence of the two-gluon-exchange
mechanism. While an energy-independent scale $\Lambda^2_{sat} = {\rm const}$
a priori cannot be strictly excluded, it appears theoretically unlikely. Among
other things, constancy would mean that the effective gluon transverse
momentum from (10) would be energy independent, the diffractively
produced $q \bar q$ mass spectrum be energy independent, the full $W$
dependence reduced to a factorizing $W$ dependence due to 
a potential (weak) energy dependence of $\sigma^{(\infty)}$
alone, etc. The generic two-gluon-exchange structure ``almost'' rules
out $\Lambda^2_{sat} = {\rm const}$ and accordingly requires saturation.

Taking advantage of the fact that the ($J = 1$ part of the) dipole
cross section (8) is essentially determined by the quantities 
$\sigma^{(\infty)}$ and $\Lambda^2_{sat} (W^2)$ in 
(9) and (10), the
total photoabsorption cross section becomes 
approximately\cite{Diff2000,Cvetic,Ku-Schi}
\begin{eqnarray}
 \sigma_{\gamma^*p} &&\hspace*{-0.5cm}(W^2, Q^2) \simeq 
\frac{\alpha R_{e^+e^-}}{3 \pi} \sigma^{(\infty)} \\
&&\begin{cases}
ln \frac{\Lambda^2_{sat} (W^2)}{Q^2 + m^2_0}, 
& $$(Q^2 << \Lambda^2_{sat} (W^2)),$$\\
\frac{1}{2} \frac{\Lambda^2_{sat} (W^2)}{Q^2} 
& $$ (Q^2 >> \Lambda^2_{sat} (W^2)).$$
\end{cases}\nonumber
\label{(11)}
\end{eqnarray}

A detailed evaluation leads to the theoretical results displayed in figs.
1 and 2.

Several remarks are appropriate:\\
i) Unitarity for the hadronic $(q \bar q)$ proton interaction
requires the integral (9) to exist and $\sigma^{(\infty)}$ to be
at most weakly dependent on the energy $W$. The fit yields
$\sigma^{(\infty)} \simeq {\rm const}~ \simeq 30 mb.$\\
ii) The existence of a scale, $\Lambda^2_{sat} (W^2)$, 
according to (10) appears as a 
straightforward consequence of the two-gluon-exchange structure. This structure
implies that the forward-scattering amplitude depends on the effective gluon
transverse momentum.\\
iii) Since unitarity $(\sigma^{(\infty)} \simeq {\rm const})$
cannot be disputed, and the assumed two-gluon exchange generic structure
seems safe, once $\Lambda^2_{sat} = {\rm const}$ is abandoned, we must
have the transition to the logarithmic behavior in (11), i.e. saturation
as depicted in fig. 2 even far beyond the energy range accessible at
present.\\ 
iv) The gluon structure function from (8) is 
given by\cite{Zakharov} $\alpha_s
(Q^2) x g (x, Q^2) = \frac{1}{8 \pi^2} \sigma^{(\infty)} \Lambda^2_{sat}
\left(\frac{Q^2}{x}\right),$ again disfavoring constancy of $\Lambda^2_{sat}
(W^2)$.\\ 
v) When saturation and the logarithmic behavior in (11) set in, the
usual connection between $F_2$ and the gluon structure function breaks
down. An extensive literature (compare e.g.\cite{Iancu} and the references
therein) attempts to apply (nonlinear) evolution equations for gluon 
distributions even in this logarithmic domain. 
\vspace*{0.1cm}

\noindent{\bf ACKNOWLEDGEMENT}\\
It is a pleasure to thank Masaaki Kuroda for a fruitful collaboration
and Roberto Fiore and Alessandro Papa for organising a very
successful meeting in splendid surroundings.


\vspace*{-0.2cm}


\begin{thebibliography}{9}
\vspace*{-0.2cm}
\bibitem{Diff2000} D. Schildknecht, in Diffraction 2000, Nucl. Phys. (Proc.
                   Suppl.) 99 (2001) 121;
                   D. Schildknecht, B. Surrow, M. Tentyukov, 
                   Phys. Lett. B499 (2001) 116.
\bibitem{Cvetic} G. Cvetic, D. Schildknecht, B. Surrow, M. Tentyukov, Eur.
                 Phys. J. C20 (2001) 77;
                 D. Schildknecht, B. Surrow, M. Tentyukov, Mod. Phys. Lett. A
                 Vol. 16 (2001) 1829;
                 D. Schildknecht, in ``The 9th International Workshop on
                 Deep Inelastic Scattering DIS 2001'', Bologna, Italy,
                 editors G. Bruni et al. (World Scientific 2002) 798.
\bibitem{Schi-Ku} M. Kuroda, D. Schildknecht, in preparation.
\bibitem{Ku-Schi} M. Kuroda, D. Schildknecht, Phys. Rev. D66 (2002)
                  094005.
\bibitem{Sakurai} J.J. Sakurai, D. Schildknecht, Phys. Lett. 40B (1972) 121;
                  B. Gorczyca, D. Schildknecht, Phys. Lett. 47B (1973) 71
\bibitem{Fraas} H. Fraas, B.J. Read, D. Schildknecht, Nucl. Phys. B86 (1975)
                346; Nucl. Phys. B88 (1975) 301;
                R. Devenish, D. Schildknecht, Phys. Rev. D19 (1976) 93.
\bibitem{Golec} A.M. Stasto, K. Golec-Biernat, J. Kwiecinski, Phys. Rev. Lett.
                B86 (2001) 596.
\bibitem{Low} F.E. Low, Phys. Rev. D12 (1975) 163; 
              S. Nussinov, Phys. Rev. Lett. 34 (1975) 1286; Phys. Rev. D14
              (1976) 246;
              J. Gunion, D. Soper, Phys. Rev. D15 (1977) 2617.
\bibitem{Nikolaev} N.N. Nikolaev, B.G. Zakharov, Z. Phys. C49 (1991) 607;
                   Z. Phys. C53 (1992) 331; Soviet Phys. JETP 78 (1994) 598.
\bibitem{Zakharov} N.N. Nikolaev, B.G. Zakharov, Phys. Lett. B332 (1994) 184.
\bibitem{Iancu} E. Iancu, K. Itakura, S. Munier, hep-ph/0310338.
                  
\end{thebibliography}
\end{document}